# More than Words: Twitter Chatter and Financial Market Sentiment[*]

Travis Adams, Andrea Ajello, Diego Silva, and Francisco Vazquez-Grande[†]

Federal Reserve Board

May 2023

**Abstract**

We build a new measure of credit and financial market sentiment using Natural Language Processing on Twitter data. We find that the Twitter Financial Sentiment Index (TFSI) correlates highly with corporate bond spreads and other price- and survey-based measures of financial conditions. We document that overnight Twitter financial sentiment helps predict next day stock market returns. Most notably, we show that the index contains information that helps forecast changes in the U.S. monetary policy stance: a deterioration in Twitter financial sentiment the day ahead of an FOMC statement release predicts the size of restrictive monetary policy shocks. Finally, we document that sentiment worsens in response to an unexpected tightening of monetary policy.

[*]The views presented here are solely those of the authors and do not represent those of the Board of Governors or any entities connected to the Federal Reserve System.

[†]Corresponding author: Francisco Vazquez-Grande, francisco.vazquez-grande@frb.gov. We wish to thank Ilknur Zer-Boudet for kindly sharing data, insights and knowledge on financial dictionaries, and feedback through the different stages of the project. We thank Craig Chikis and Tyler Pike for their pioneering efforts in pursuing this analysis. We are grateful to John Schindler and Nicholas Wolanske for their outstanding research assistance. We thank Louisa Kontoghiorghes (discussant), Steve Sharpe, Tomaz Cajner, Giovanni Favara, Don Kim, Seung Lee, Francesca Loria, Ander Perez-Orive, Yannick Timmer and participants in the ECONDAT2023 Bank of England and ECB joint conference, the Fed Board Hackathon events, the Macro-Financial Analysis and Twitter Workgroup brown bag seminars for comments and suggestions. All errors remain our own.

# 1   Introduction

Does social media activity carry any meaningful signal on credit and financial markets' sentiment? We build a new real-time sentiment index derived from social media communications related to credit and financial markets. We rely on sentiment analysis of Twitter data and show that financial sentiment gauged from social media contains predictive information for stock returns and proves sensitive to monetary policy surprises, predicting tightening moves ahead of FOMC statement releases, as measured by several event-study monetary policy shocks developed in the literature.

We query a large sample of tweets that contain words and word clusters from financial- and credit-market dictionaries (Calomiris and Mamaysky, 2019), from the universe of social media posts available on Twitter since 2007. For each tweet in our sample, we measure sentiment using FinBERT a language model developed by Araci (2019) from BERT (Devlin et al., 2018) and specifically designed to measure sentiment of financial text. Our index draws from the universe of Twitter users who post financial content and is available in real time, as new tweets appear on the platform and their sentiment is assessed. Averaging sentiment values of posted tweets, we build a historical index of financial market sentiment and name it the Twitter Financial Sentiment Index (TFSI). We document that time variation in the TFSI can be attributed to changes in the extensive margin of users engaging in posting positive or negative sentiment tweets, rather than to the intensive margin—i.e., users posting tweets with higher or lower sentiment.

We show that the monthly TFSI correlates highly with market-based measures of financial sentiment, such as corporate bond spreads, the Excess Bond Premium (EBP) (Gilchrist and Zakrajšek, 2012), and survey-based measures of consumer confidence, such us the Michigan confidence index. We also find that our index correlates positively with market-based measures of borrowing costs, such as corporate credit spreads.

With the index at hand, we make two main contributions. First, we show that overnight Twitter sentiment can help predict daily stock market returns–i.e., the average tweeted sentiment between 4pm on day $t-1$ to 9am on day $t$ helps forecast open-to-close stock market returns on day $t$ after controlling for standard asset pricing factors. This fact speaks to the ability of tweeted sentiment to reflect information that will later be included in stock prices once U.S. markets open.

Second, the TFSI predicts the size of restrictive monetary policy surprises. We show that Fed-related tweets play a dominant role on FOMC days and, notably, that Twitter sentiment after the first day of the FOMC meeting can predict the size of restrictive monetary policy shocks in connection with the release of the FOMC statement the following day. This last results holds across three measures of monetary policy shocks, identified by means of event studies in Miranda-Agrippino and Ricco (2021), Jarociński and Karadi (2020), and Bauer and Swanson (2022). In other words, Twitter financial sentiment ahead of monetary policy decisions incorporates useful information that can help predict the market reaction around the FOMC statement release. We also find that the TFSI further worsens in response to an unexpected tightening in the policy stance, but does not respond systematically to easing shocks.

We contribute to the literature that attempts to measure financial market sentiment (see for example López-Salido et al. (2017); Danielsson et al. (2020), Greenwood and Hanson (2013), Shiller (2015b), Fama and French (1988), Baker and Wurgler (2000) and Lettau and Ludvigson (2001)), employing natural language processing to harness information from Twitter posts as a novel data source.

Time variation in average sentiment across tweets can broadly capture changes in expectations, risk appetite, beliefs, or emotions representative of a wide array of Twitter users. Traditional gauges of financial market sentiment are based on asset prices, portfolio allocation flows (Baker and Wurgler, 2006; Gilchrist and Zakrajšek, 2012), investors' surveys (Qiu and Welch, 2004), and news archives (Tetlock, 2007; Garcia, 2013). While measures based on portfolio allocations, prices, and news coverage can be monitored at high frequency, survey measures imply that sentiment is polled infrequently. Such sentiment measures are derived from actions, market outcomes, opinions and commentary from selected groups of actors rather than from the wider public.

Time variation in credit and financial market sentiment has proven to be an important predictor of asset returns (Shiller, 2015a; Greenwood and Hanson, 2013), and driver of credit and business cycles (López-Salido et al., 2017), and we aim to explore this transmission by means of our index in the near future and in future iterations of this paper.

Moreover, central bank decisions and communication strategies, intended to fine-tune the stance of monetary policy and share information on the state of the central bank's economic outlook, affect market participants' expectations, risk sentiment, and beliefs, as policy transmits to the broader economy (Gertler and Karadi, 2015; Miranda-Agrippino and Rey, 2020; Bekaert et al., 2013).

Our paper relates to a particular strand of literature that studies the role of text-based measure of financial market sentiment. Financial sentiment measured from news archives has been shown to predict stock market performance (Tetlock, 2007; Garcia, 2013). Research based on social media data show that the Twitter activity of institutions, experts, and politicians contains useful information to study various aspects related to central banking. As central banks have become more active on Twitter (Korhonen and Newby, 2019; Conti-Brown and Feinstein, 2020), Azar and Lo (2016) find that tweets that refer to FOMC communication can help predict stock market returns. Our results show that twitter sentiment can help predict stock market returns more systematically and can anticipates changes in the stance of monetary policy. Masciandaro, Peia, and Romelli (Masciandaro et al.) use dissimilarity between Fed-related tweets and FOMC statements to identify monetary policy shocks, while Meinusch and Tillmann (2017), Stiefel and Vivès (2019), and Lüdering and Tillmann (2020) use tweets to estimate changes in public beliefs about monetary policy and their impact on asset prices, although they do not explore the ability of twitter sentiment to forecast monetary policy shocks. Ehrmann and Wabitsch (2022) focus on studying view divergence and polarization in response to central bank communication. They show that following ECB communication, tweets primarily relay information and become more factual and that public views become more moderate and homogeneous. High-impact decisions and communications, such as Mario Draghi's "Whatever it takes" statement, instead trigger a divergence in views.

Recent work applies sentiment analysis to a wider set of central bank communication tools. Notably, Correa et al. (2020) measure sentiment in central banks' financial stability reports, introducing a dictionary tailored to financial stability communications, confirming that general dictionaries, including finance dictionaries such as Loughran and McDonald (2016), might not be suitable to assess tonality in a financial stability context. Binder (2021), Bianchi et al. (2019), Camous and Matveev (2021), and Tillmann (2020) show that tweets by former U.S. president Trump about the Federal Reserve and its policy stance affected long-term inflation expectations and confidence of consumers, suggesting that the wider public priced in future reductions in interest rates in response to the president's social media activity. Finally, Angelico et al. (2022) show that Twitter can be an informative data source to elicit inflation expectations in real time.

# 2 Methodology

In this section we describe our strategy to sample financial tweets from the Twitter historical and real-time enterprise-level Application Programming Interface (API) (Twitter, Inc., Inc.). We then describe how we filter the data, pre-process tweets, and compute sentiment to produce readings of our index at different time frequencies.

## 2.1 Sampling

We query a subset of tweets related to financial market developments from the universe of all tweets available since 2007. Calomiris and Mamaysky analyze news articles from the Thompson Reuters archive and isolate a set of 60 word roots related to financial discourse that we use to discipline the sample selection of historical and real-time tweets, downloaded from the Twitter APIs.

Downloading all tweets that contain any combination of word roots in the Calomiris and Mamaysky set proves undesirable and infeasible: word derived from roots in the set can have multiple meanings–e.g., the word "bond" can be used to mean "connection" as well as "fixed income obligation". A large, unsystematic query has a higher likelihood of contaminating the sample with non-financial tweets—and surpasses by at least one order of magnitude our contracted Twitter API download quota.

To discipline the sample of tweets, we use Keyword Clustering, pairing the set of word roots into groups that are semantically similar. We measure similarity across keywords by means of their cosine distances from machine-learning-generated semantic similarity vectors (Yamada et al., 2020): a trained machine assesses the similarity of our keywords based on their occurrence within the body of text of Wikipedia. Figure 1 shows that the three clusters loosely map into financial contracts (Group 1), entities (Group 2), and actions or contractual features (Group 3).

Our query uses logical operators to filter tweets that contain at least one word from each of the three clusters. Technical features of the Twitter API require that single tweets be downloaded as separate entities—that is, the search engine treats threads and quote tweets as disconnected tweets, while retweets are linked to the original tweet by means of a boolean operator and share their creating time and date with the original tweet even when they were posted at a later time. We pre-process the text of all tweets by removing excess white spaces, tags, hyperlinks, and information that is not part of the text body of the tweet. We only keep tweets with unique text, filtering out full and near replicas of tweets, to reduce the number of bot-generated entries in our dataset.[1]

Figure 1: Semantic Similarity Clusters

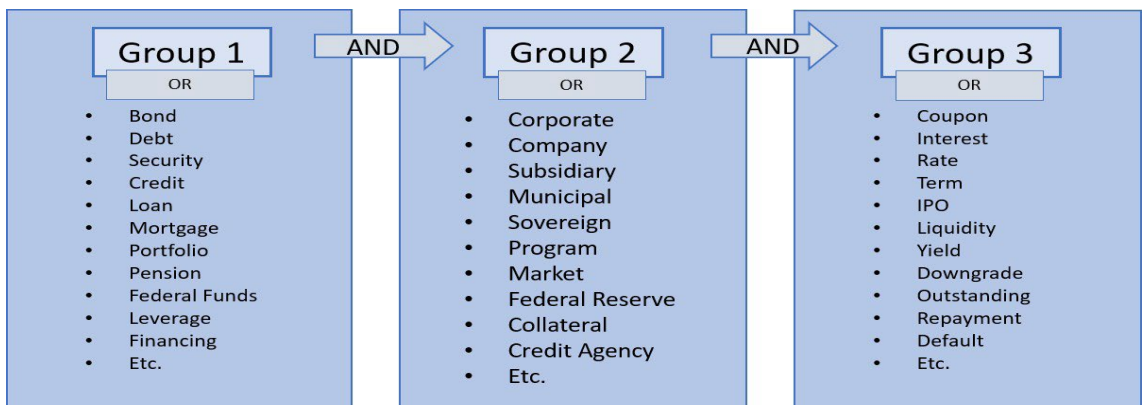

Note: This diagram displays an example of how financial words are clustered in three groups based on semantic similarity. See Appendix B for the full list of word and word roots, by cluster.

Our data query and preprocessing deliver a total of 4.4 million single tweets from 2007 to April 2023. Figure 2, plots the number of tweets downloaded per month since the beginning of the sample. Two structural features affect our data query. First, prior to 2011, as Twitter was burgeoning as a social media platform and its popularity was low, the number of tweet pulls averages around one hundred tweets per day, offering limited amount of text to measure sentiment at daily or weekly frequency. Second, in November 2017 Twitter increased the maximum character length of tweets from 140 to 280 characters, a change that makes it more likely to detect any three-word sequence in our query within a single post, resulting in a discrete jump in tweets pulled each month thereafter. Worth of note is the fact that discrete events, such as the start of the COVID-19 pandemic, the pivot in communication toward a tightening cycle in September 2021, and more recently the collapse of Silicon Valley Bank positively affect the number of tweets in the data pull from our query. The sample we use for our baseline analysis in sections 3 and 4 starts in September 2011—after the step increase in the number of monthly pulls visible in Figure 2—and includes 4.3 million tweets.

---

[1]In a similar spirit, we filter out tweets that advertise credit cards, crypto currency trades, and tweets related to topics that are only seemingly related to financial or credit market discourse, such as those that include words like "social security". Appendix B contains the full list of words from Calomiris and Mamaysky clustered in the three groups, and a detailed methodology to replicate our tweet selection and data cleaning.

Figure 2: Number of Tweets Selected per Month

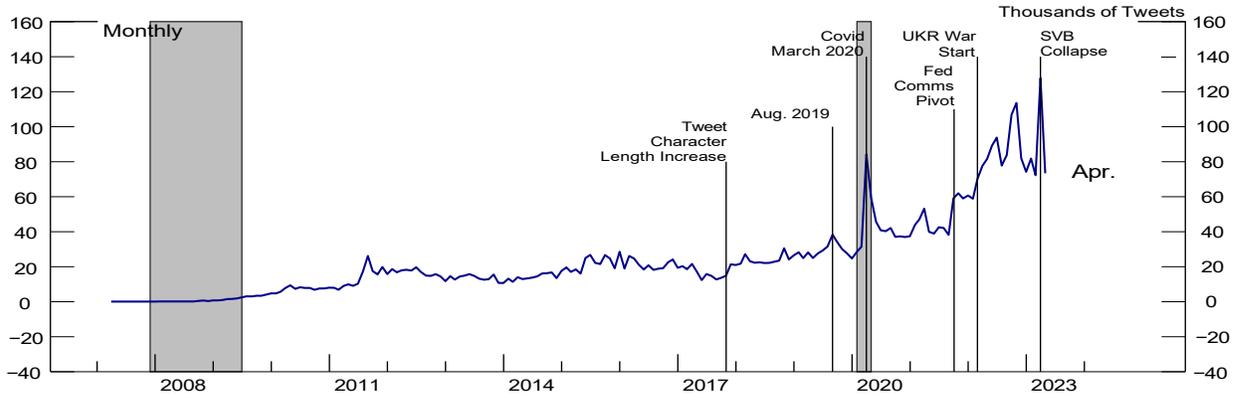

Note: This figure represents the number of tweets in the sample per month. The shaded bars indicate periods of business recession as defined by the National Bureau of Economic Research: December 2007–June 2009, February 2020–April 2020.
Source: Authors' calculation based on Twitter enterprise-level API data.

## 2.2 Measuring Sentiment

We use FinBERT (Araci, 2019) as our baseline tool to compute a sentiment value for each tweet of our sample. FinBERT is a language model based on Bidirectional Encoder Representations from Transformers (BERT) (Devlin et al., 2018) to tackle natural language processing tasks in financial domain. An advantage of this tool relative to other text sentiment gauges is that it is specifically designed and trained to perform well measuring sentiment of financial text, making it the ideal candidate for our purpose.

We use FinBERT's compound score to assign a numeric value of sentiment to each tweet. FinBERT provides three probabilities to measure the odds that the analyzed text conveys positive neutral or negative sentiment, and also offers a compound sentiment score computed as the difference between the probability of the text having positive sentiment and the probability of the text having negative sentiment. Therefore FinBERT, provides us with a sentiment score between -1 and +1 for each tweet in our sample. For the purpose of measuring the evolution of average sentiment over time, given a time window, we first assign a sentiment score of zero to tweets that are labeled as neutral, and then sum the sentiment score of all remaining tweets and divide by the total number of tweets posted over the desired time frame. Following this methodology, we can average sentiment across all tweets sampled in any desired time span and compute financial sentiment values at different time frequencies.[2]

## 3   The Twitter Financial Sentiment Index

The TFSI is calculated as the average sentiment across tweets in our sample for any given time period, as such it can be displayed in real time and at any time frequency. Figure 3, plots the TFSI at daily and monthly frequencies (top and bottom panel), since 2011. The daily index is particularly volatile at high frequencies especially before the Twitter CLI in 2017, after which the sentiment signal appears more informative day-to-day. For ease of comparability with other gauges of financial conditions, the index is oriented so that higher values indicate a deterioration in sentiment: the index rises in alignment with episodes of elevated stress in the U.S. financial system or tightening of financial conditions. These episodes include the Taper Tantrum in 2013, market selloff related to emerging market stress in 2014, and turbulence in late 2015 and early 2016 associated with fears related to the Chinese economy and a pronounced drop in oil price, the increase in likelihood that the US economy would enter a recession in August 2019, the COVID recession in 2020, and in the souring of financial sentiment associated with the onset of the Russian conflict in Ukraine and the beginning of tightening of monetary policy with the Fed communication pivot in September 2021.

---

[2]For robustness, we also measure sentiment using VADER, a lexicon and rule- based sentiment analysis tool that is specifically designed to measure sentiment expressed in social media (see Hutto and Gilbert (2014)). An advantage of this tool relative to other text sentiment gauges is that is better equipped to parse modifier words, and emojis to assess sentiment in social media text. Results using VADER sentiment are available upon request.

Figure 3: Twitter Financial Sentiment Index, Daily (Up), Monthly (Down)

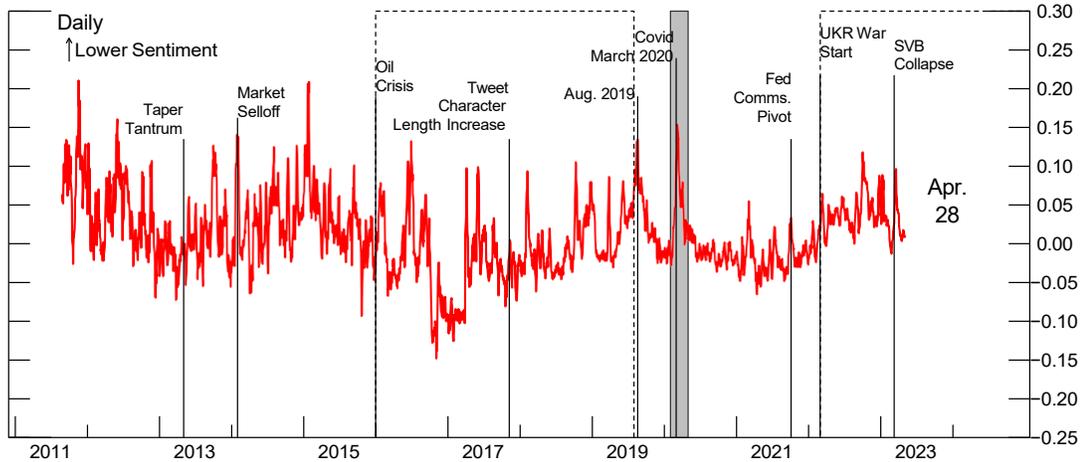

(a) Daily Twitter Financial Sentiment Index — Seven-Day Moving Average

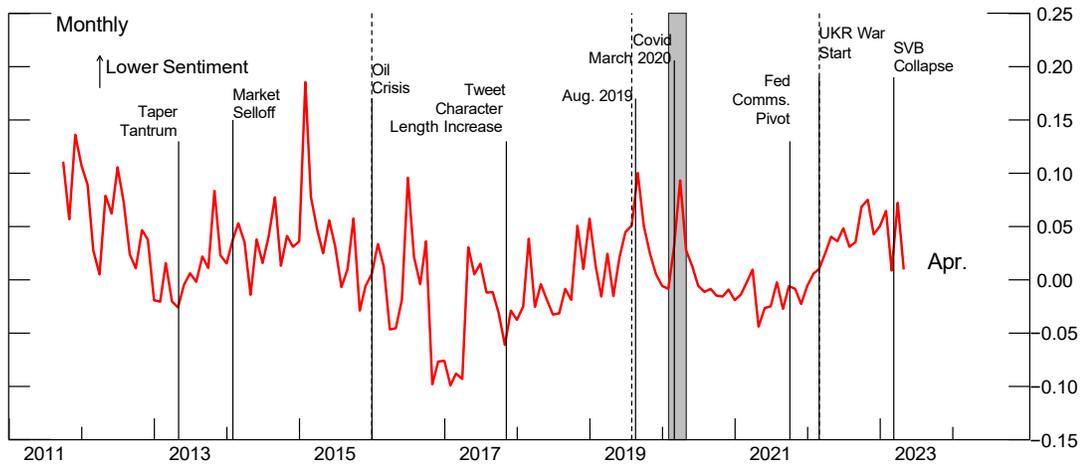

(b) Monthly Twitter Financial Sentiment Index

Note: Increases in the TFSI point to a worsening of sentiment. Data sample starts from September 2011 and ends in April 2023. The dashed boxes indicate monetary policy tightening cycles: January 2016- August 2019, March 2022-present. The shaded bar indicate periods of business recession as defined by the National Bureau of Economic Research: February 2020–April 2020.
Source: Authors' calculation based on Twitter enterprise-level API data.

We also find that time-variation in the index can be mostly explained by the share of users that post tweets with positive or negative sentiment, rather than by the intensity of the tweeted sentiment. In principle, the value of the index would vary by changes in the intensive or the extensive margin, that is, it could be driven by changes in the sentiment value of the tweets or by the share of tweets with positive or negative sentiment values posted in a given time interval. Figure 4 compares our baseline index (in red) with the share of negative minus the share of positive tweets (in green), a measure of engagement on either side of the sentiment fronts. The similarity between the two lines demonstrates how most of the variation in the index is related to users' engagement on the extensive margin, rather than in the intensity of their sentiment expressed in their tweets.

Figure 4: Twitter Financial Sentiment Index vs. Share of Negative minus Share of Positive Tweets

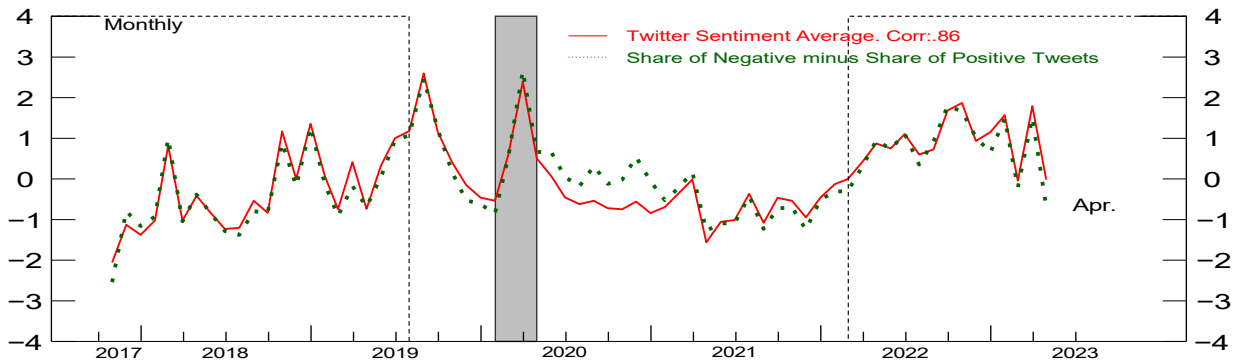

Note: The chart plots the Twitter Financial Sentiment Index (red solid line), against the difference in share of negative- and positive-sentiment tweets (green dashed line) at a monthly frequency and both standardized. Data sample starts from September 2011 and ends in April 2023. Increases in the TFSI point to a worsening of sentiment. The dashed boxes indicate monetary policy tightening cycles: January 2016- August 2019, March 2022-present. The shaded bar indicate periods of business recession as defined by the National Bureau of Economic Research: February 2020–April 2020.
Source: Authors' calculation based on Twitter enterprise-level API data.

# 4   Results

This section summarizes our main results. We show that the TFSI correlates with indexes and market gauges of financial conditions at monthly frequency. We also show that overnight twitter sentiment can help predict daily stock market returns. Finally, we show that Twitter financial sentiment can predict the size of restrictive monetary policy surprises and has a muted response to the realization of monetary policy shocks.

## 4.1   TFSI and Financial Conditions

Figure 5, compares the monthly TFSI with measures of financial conditions and economic and financial sentiment based on surveys and market prices since the Twitter CLI: the Baa corporate bond spread (top), and the Excess Bond Premium (EBP) of Gilchrist and Zakrajšek (2012) (middle) and the University of Michigan Consumer Sentiment index (bottom). The TFSI, while noisier, generally co-moves positively with these measures. These figure show that our sample selection and sentiment measure, that does not depend at all on market prices or surveys, presents a quantitatively and qualitatively similar picture to the most common metrics of economic and financial conditions.

Figure 5: TFSI vs Measures of Financial Conditions and Sentiment

TFSI and Baa Corporate Bond Spreads

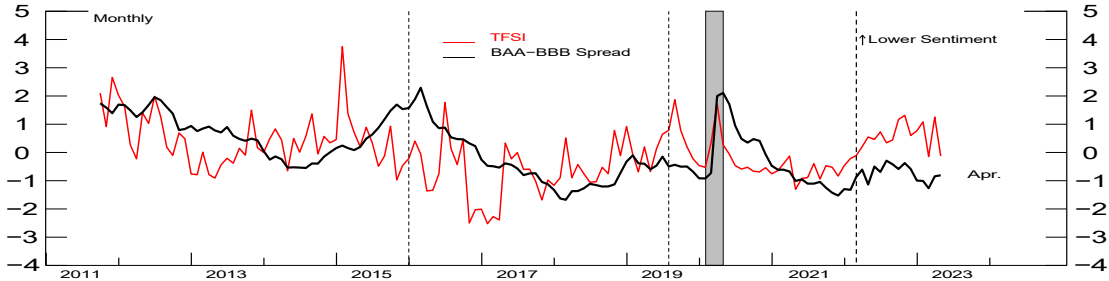

TFSI and Excess Bond Premium

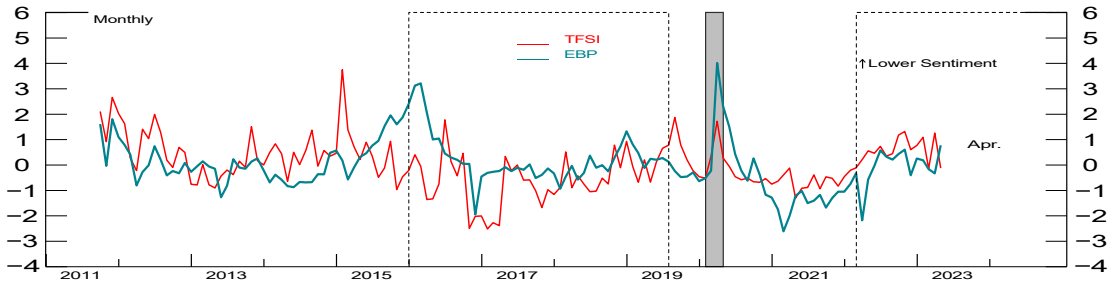

TFSI and U. Michigan Consumer Sentiment Index

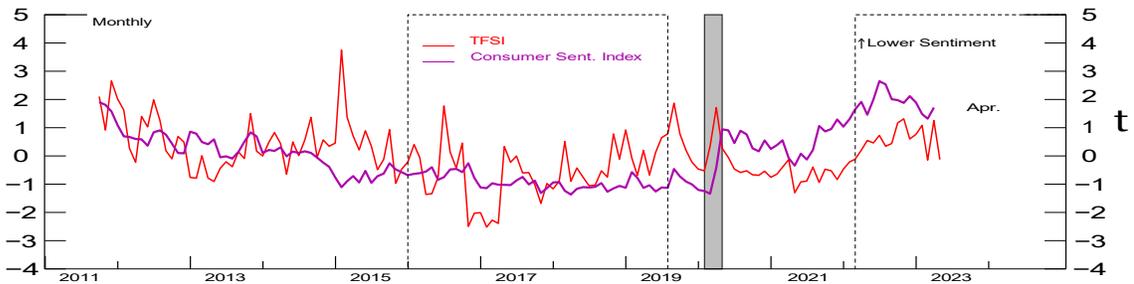

Note: Increases in the TFSI point to a worsening of sentiment. The dashed boxes indicate periods of monetary policy tightening cycles: January 2016-August 2019, March 2022-present. The shaded bars indicate periods of business recession as defined by the National Bureau of Economic Research: February 2020–April 2020.
Source: TFSI: Authors' calculation based on Twitter Enterprise-level API; Baa Spreads: Moody's via FRED; EBP: Federal Reserve Board, Favara et al. (2016); Consumer Sentiment Index: University of Michigan via FRED

## 4.2 TFSI and Stock Market Returns

We show that the TFSI can be used to forecast intraday returns of the S&P 500 index, even after controlling for other common predictors such as the VIX, the Fama-French stock market factors (Fama and French, 2015), financial sentiment present in official media sources (Shapiro, 2020), and lagged S&P 500 returns. One advantage of twitter data is that it is available in real time, 24 hours a day. We take advantage of this feature to construct a measure of sentiment available when financial markets are closed. We compute the sentiment of all tweets in our sample that are posted overnight, that is, between 4pm at date $t-1$ and 9am at date $t$. We then run the following daily regressions:

$$SP500_{t_{(9am)} \to t_{(4pm)}} = \alpha + \beta \ TFSI_{t-1_{(4pm)} \to t_{(9am)}} + \gamma \ X_{t-1} + \varepsilon_t$$

where $SP500_{t_{(9am)} \to t_{(4pm)}}$ are the daily intraday Standard and Poor's 500 index market returns (from S&P Global, CapitalIQ), $TFSI_{t-1_{(4pm)} \to t_{(9am)}}$ is our measure of overnight sentiment, and $X_{t-1}$ is a vector of controls.

Table 1 displays the results. Each column of the table adds common predictors of daily returns as controls, that is, lagged S&P 500 index returns, the overnight return on the S&P 500—the returns between the close of market on day $t-1$ and opening of the market on day $t$—the VIX, and the three stock market factors of Fama and French (HML, High minus Low, SMB, Small minus Big, and MOM, Momentum). In all specifications the overnight sentiment index has a negative and significant coefficient. Each column also adds controls for financial sentiment in newspaper articles to account for sentiment in conventional media as measured by Shapiro (2020). All stock market variables are expressed as daily rates of return, while sentiment indexes are expressed in standard deviations. All variables are stationary. We find that lower sentiment overnight predicts lower stock returns the following business day. In terms of magnitude a one-standard-deviation increase overnight in the TFSI, everything else equal, leads to a decrease of about 6 basis point in daily S&P 500 index returns. In Appendix A.8 we build and backtest a trading strategy that conditions long or short trades on the S&P 500 index on a threshold value for overnight TFSI (e.g., buy at open and sell at close if overnight sentiment is positive and vice versa if sentiment is negative). We find that such strategy outperforms a simple benchmark that goes long daily on the S&P 500.

The TFSI also correlates contemporaneously with stock returns. Table 2 presents the results of regressing daily S&P 500 index returns on the contemporaneous observation of the TFSI (measured between 4pm at date $t-1$ and 4pm at date $t$) and the same control variables as in Table 1, excluding overnight returns. The contemporaneous TFSI also displays a negative and significant coefficient which implies that the worse the sentiment, as measured by the TFSI, the lower the daily S&P 500 index returns. In terms of magnitude a one-standard-deviation increase in the TFSI, everything else equal, corresponds to a decrease of about 10 basis point in daily aggregate returns. We do not find a statistically significant relation to aggregate market returns using one-day-lagged TFSI as a regressor (results not shown).[3]

---

[3] We test the assumption that the residuals of all models in tables 1 and 2 are i.i.d. (White, 1980), and we find that the assumption is rejected for all models, excluded model (2) that controls for news sentiment. To account for the role of heteroskedasticity on the uncertainty around the models' estimated coefficients, we report HAC-robust standard errors (Andrews, 1991).

Table 1

|  | Dependent variable: |  |  |  |  |
| --- | --- | --- | --- | --- | --- |
|  | $SP500_{t_{(9am)} \to t_{(4pm)}}$ |  |  |  |  |
|  | (1) | (2) | (3) | (4) | (5) |
| $TFSI_{t-1_{(4pm) \to t_{(9am)}}}$ | -0.05** | -0.06*** | -0.06*** | -0.06*** | -0.06*** |
|  | (0.02) | (0.02) | (0.02) | (0.02) | (0.02) |
| $News_t$ |  | -0.24** | 0.01 | 0.01 | 0.01 |
|  |  | (0.10) | (0.15) | (0.15) | (0.15) |
| $FOMC_t$ |  |  |  | 0.06 | 0.07 |
|  |  |  |  | (0.08) | (0.08) |
| $HML_{t-1}$ |  |  |  |  | −0.04 |
|  |  |  |  |  | (0.03) |
| $SMB_{t-1}$ |  |  |  |  | -0.004 |
|  |  |  |  |  | (0.04) |
| $MOM_{t-1}$ |  |  |  |  | −0.06*** |
|  |  |  |  |  | (0.02) |
| $SP500_{t-1}$ |  |  | -0.04* | -0.04* | −0.05** |
|  |  |  | (0.03) | (0.03) | (0.03) |
| $SP500_{t-1_{(4pm) \to t_{(9am)}}}$ |  |  | 0.52*** | 0.52*** | 0.52*** |
|  |  |  | (0.07) | (0.07) | (0.07) |
| $VIX_{t-1}$ |  |  | 0.01 | 0.01 | 0.01 |
|  |  |  | (0.006) | (0.006) | (0.006) |
| Constant | 0.03** | 0.03** | -0.11 | -0.11 | −0.11 |
|  | (0.02) | (0.02) | (0.10) | (0.10) | (0.10) |
| Observations | 2,956 | 2,955 | 2,955 | 2,955 | 2,955 |
| Adjusted R² | 0.002 | 0.004 | 0.08 | 0.08 | 0.08 |

*Note:* *p<0.1; **p<0.05; ***p<0.01

This table regresses returns of the S&P 500 index on a twitter based measure of "overnight" sentiment and an expanding set of controls used in the literature to forecast stock market returns:

$$SP500_{t_{(9am)} \to t_{(4pm)}} = a + \beta \ TFSI_{t-1_{(4pm) \to t_{(9am)}}} + \gamma \ X_t + \varepsilon_t$$

$News_t$ represents the sentiment in official news sources as calculated by Shapiro (2020). $SP500_{t-1}$ are the daily Standard and Poor's 500 index market returns. $TFSI_{t-1(4pm) \to t(9am)}$ is our sentiment index from $4pm_{t-1}$ to $9am_t$. $FOMC_t$ is a binary variable indicating if the day in question was an FOMC meeting day. The variables $HML_t$, and $SMB_t$, represent the High-minus-low and Small-minus-big Fama-French factors (1993) respectively. The variable $MOM_t$ represent the momentum factor as defined by Cahart (1997). $SP500_{t-1}$ denotes Standard and Poor's 500 index market returns from the previous day. $SP500_{t-1(4pm) \to t(9am)}$ denotes Standard and Poor's 500 index market returns from the close of the market in the previous day to the opening of the market in the current day. $VIX_t$ represent the implied volatility index from the Chicago Board Options Exchange. The sample goes from September 2011 until April 2023. The table reports HAC-robust standard errors for all coefficient estimates (in parentheses).

## 4.3 TFSI and Monetary Policy

We find that the tweets in our sample relate strongly to federal reserve communications in and around FOMC days. Figure 6 shows two word clouds obtained from tweets in our sample. In such diagrams, the size of the words displayed is proportional to the word's frequency in the body of text. On the left we show the word cloud across all the tweets in our sample, and on the right the word cloud on FOMC days. Words associated with Federal Reserve communication are clearly displayed more prominently in the FOMC-days-only word cloud, which suggests that the twitter discourse in our sample on FOMC days is driven by monetary policy decisions.

Figure 6: Frequent Words: All Sample and on FOMC Days

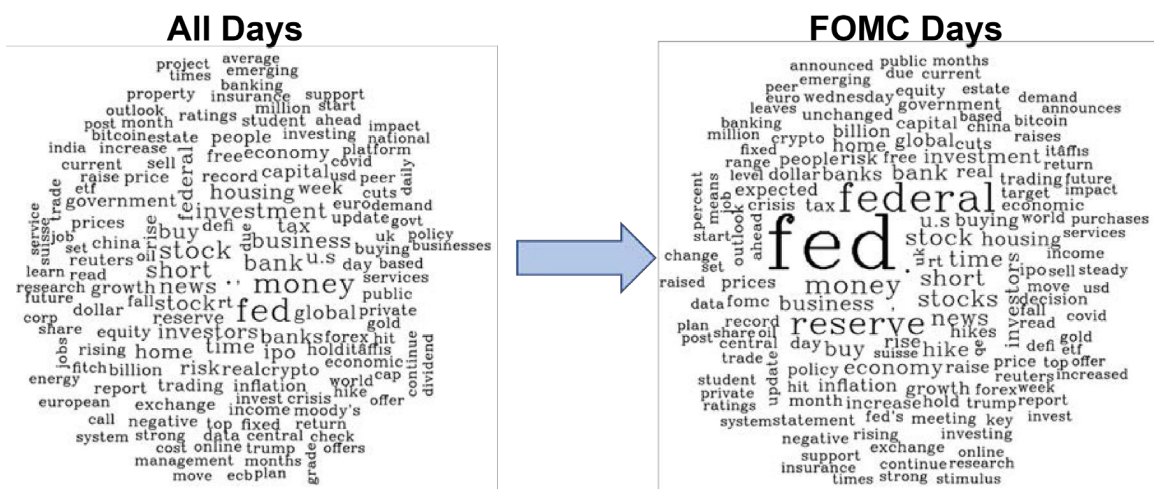

It is also worth noting that in proximity of an FOMC meeting the prevalence of tweets related to the Fed and to monetary policy increases. Figure 7 plots the average share of Fed-related tweets in our sample against the number of calendar days away from the second day of the FOMC meeting. On FOMC days the share of fed related tweets is about 25 percent on average. The share remains significantly above average, between one day before and 5 days after the FOMC meeting, reverting back close to its sample mean of 12 percent (the dashed line).

With these observations at hand, we study how Twitter sentiment behaves ahead and after monetary policy decisions. We find that the TFSI helps predict the size of restrictive monetary policy surprises, while it is uninformative on the size of easing shocks.

Twitter sentiment measured ahead of the release of the official monetary policy determina-

Table 2

|  | Dependent variable: | | | | |
|---|---|---|---|---|---|
|  | $SP500_t$ | | | | |
|  | (1) | (2) | (3) | (4) | (5) |
| $TFSI_t$ | -0.10*** | -0.12*** | -0.10*** | -0.10*** | -0.09*** |
|  | (0.02) | (0.02) | (0.03) | (0.03) | (0.03) |
| $News_t$ |  | -0.46*** | -1.66*** | -1.66*** | -1.58*** |
|  |  | (0.14) | (0.24) | (0.24) | (0.24) |
| $FOMC_t$ |  |  |  | 0.15 | 0.17 |
|  |  |  |  | (0.11) | (0.11) |
| $HML_t$ |  |  |  |  | -0.08* |
|  |  |  |  |  | (0.05) |
| $SMB_t$ |  |  |  |  | 0.21*** |
|  |  |  |  |  | (0.05) |
| $MOM_t$ |  |  |  |  | -0.19*** |
|  |  |  |  |  | (0.03) |
| $SP500_{t-1}$ |  |  | -0.17** | -0.17*** | -0.18** |
|  |  |  | (0.07) | (0.07) | (0.02) |
| $VIX_t$ |  |  | -0.05*** | -0.05*** | -0.04*** |
|  |  |  | (0.010) | (0.010) | (0.010) |
| Constant | 0.06*** | 0.05** | 0.87*** | 0.86*** | 0.84*** |
|  | (0.02) | (0.02) | (0.16) | (0.16) | (0.17) |
| Observations | 2,956 | 2,955 | 2,955 | 2,955 | 2,955 |
| Adjusted $R^2$ | 0.01 | 0.04 | 0.04 | 0.09 | 0.09 |

*Note:* *p<0.1; **p<0.05; ***p<0.01

Note: This table regresses returns of the S&P 500 index on the daily TFSI (a twitter based measure of financial sentiment) and an expanding set of controls used in the literature to forecast stock market returns:

$$SP500_t = a + \beta_1 TFSI_t + \beta_2 FOMC_t + \beta_3 X_t + \beta_4 Returns_{t-1} + \beta_5 VIX_t + \varepsilon_t$$

$News_t$ represents the sentiment in official news sources as calculated by Shapiro (2020). $SP500_{t-1}$ are the daily Standard and Poor's 500 index market returns. $TFSI_t$ is the daily simple average of all unique finance-related and credit-related tweets on a scale from -1 to 1. $FOMC_t$ is a binary variable indicating if the day in question was an FOMC meeting day. The variables $HML_t$, and $SMB_t$, represent the High-minus-low and Small-minus-big Fama-French factors (1993) respectively. The variable $MOM_t$ represent the momentum factor as defined by Cahart (1997). $SP500_{t-1}$ denotes Standard and Poor's 500 index market returns from the previous day. $VIX_t$ represent the implied volatility index from the Chicago Board Options Exchange. The sample goes from September 2011 until April 2023. The table reports HAC-robust standard errors for all coefficient estimates (in parentheses).

Figure 7: Average share of Fed-related tweets against calendar days away from FOMC date.

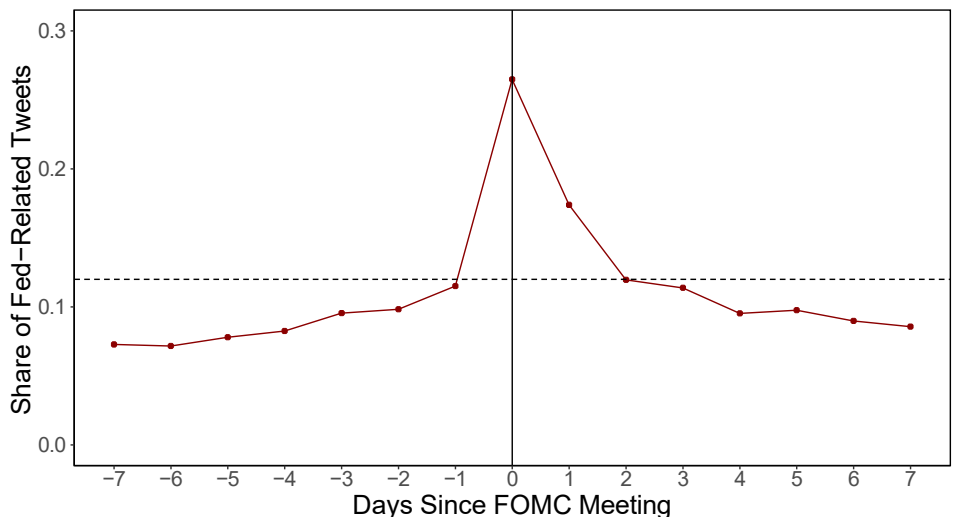

Note: The red line plots the daily share of Fed-related tweets defined in Appendix A, one week before and one week after an FOMC statement release. The dashed line represents the full sample average of Fed-related tweets standing at 12 percent.
Source: TFSI: Authors' calculation based on Twitter enterprise-level API data

tions of the Federal Open Market Committee (FOMC) can predict the size of restrictive monetary policy shocks as gauged by event-study monetary policy shocks. Our finding holds across three measures of monetary policy shocks that control for the central bank information effect—or changes in policymakers' assessment of the macroeconomic outlook conveyed by the policy statement: (Miranda-Agrippino and Ricco, 2021, henceforth MAR), (Jarociński and Karadi, 2020, henceforth JK), and (Bauer and Swanson, 2022, henceforth Bauer and Swanson).[4] Our results imply that tweeted financial sentiment ahead of monetary policy decisions contains information that can help predict the market reaction around the FOMC statement release.

Tables 3 regresses three different measures of monetary policy surprises on the TFSI index value measured over the time window between 4pm the day before the FOMC statement release and 2pm (excluded) on the day of the release. The first column of each table uses all monetary policy shocks in the sample, while the second and third columns split the sample in restrictive and easing shocks, respectively. The first columns of each table suggests that no systematic significant correlation holds between monetary policy shocks and values of the TFSI, for any of the three types of monetary policy shocks. After splitting the sample into tightening and easing shocks, however, the second columns reveal that the TFSI ahead of the policy announcement is a significant predictor of the size of restrictive monetary policy shocks, while this is not the case ahead of easing shocks. In other words, the TFSI increases (and sentiment sours) ahead of tighter monetary policy shocks.[5] Unexpected monetary policy moves—that should be unforecastable—are in fact debated in the Twitter conversation and affect its sentiment ahead of FOMC decisions. A negativity bias (Baumeister et al., 2001) seems be at play by which the anticipation of a negative outcome (a monetary policy tightening) is more likely to be reflected in Twitter sentiment relative to the anticipation of a positive outcome (a monetary policy easing).

---

[4]MAR shocks are computed from 30-minute-window changes in the 2-year on-the-run Treasury Yield around policy announcements over a sample that starts in (sample: September 2011 to December 2022). JK shocks are computed from 30-minute-window changes in the three month ahead monthly Fed Funds futures (FF4) quotes around policy announcements, limiting the sample to those episodes in which the sign of the FF4 surprise and SP500 surprise have the opposite sign (sample: September 2011 to December 2019). Bauer and Swanson's shocks are computed from 30-minute-window changes in the FF4 quotes around policy announcements orthogonalized with respect to macroeconomic and financial data that pre-date the announcement (sample: September 2011 to December 2022).

We also represent these results graphically for two out of the three series of publicly available shocks. Figure 8 plots the size of the monetary policy shocks of JK (top), and Bauer and Swanson (bottom) on the x axis against the TFSI on the y axis. As expected, we find a statistically significant relation between our measure of sentiment and the different gauges of tightening shocks. Larger contractionary monetary policy shocks are associated with souring sentiment— an increase in our measured sentiment values. Easing monetary policy shocks, however, do not elicit an improvement in sentiment.

Finally, we study whether the size of monetary policy shocks affect sentiment after the release of the FOMC statement. Models in Columns 1, 3, and 5 of Table 4 regress the TFSI after the monetary policy announcement on the full set of monetary policy shocks, tightening shocks, and easing shocks respectively. Columns 2, 4 and 6 also add the TFSI before the statement release as an additional control to the univariate models. Columns 3 of Table 4 suggest that the TFSI— measured between 2PM and 4PM on the day of the policy announcement—responds significantly to unexpected tightening in the policy stance across all three shock measures, but this effect weakens once we control for twitter sentiment measured before the policy statement release (Column 4). Columns 2, 4, and 6 suggest that sentiment ahead of the policy announcement is a

---

[5]The statistical significance of these results is preserved after controlling for financial sentiment in media, as measured by Shapiro (2020), the returns of the SP500 and the level of the VIX index. Results are available upon request.

# Table 3: TFSI and Monetary Policy Shocks — Prediction

## MAR shocks

|  | Dependent variable: | | |
|---|---|---|---|
|  | MAR Shocks$_t$ | | |
|  | All | Tight | Ease |
| $TFSI_{t-1\ (4pm) \to t}$ | 0.03 | 0.51*** | -0.06 |
|  | (0.10) | (0.12) | (0.16) |
| Constant | 0.001 | -0.00 | 0.00 |
|  | (0.10) | (0.12) | (0.16) |
| Observations | 93 | 52 | 41 |
| Adjusted R² | -0.01 | 0.24 | -0.02 |

## JK shocks

|  | Dependent variable: | | |
|---|---|---|---|
|  | JK Shocks$_t$ | | |
|  | All | Tight | Ease |
| $TFSI_{t-1(4pm) \to t(1:59pm)}$ | -0.13 | 0.58*** | -0.21 |
|  | (0.13) | (0.15) | (0.18) |
| Constant | 0.00 | -0.00 | -0.00 |
|  | (0.13) | (0.15) | (0.18) |
| Observations | 62 | 30 | 32 |
| Adjusted R² | 0.001 | 0.32 | 0.01 |

## BS shocks

|  | Dependent variable: | | |
|---|---|---|---|
|  | Bauer - Swanson Shocks$_t$ | | |
|  | All | Tight | Ease |
| $TFSI_{t-1(4pm) \to t(1:59pm)}$ | 0.07 | 0.36** | 0.005 |
|  | (0.12) | (0.16) | (0.20) |
| Constant | -0.003 | -0.00 | 0.00 |
|  | (0.13) | (0.16) | (0.19) |
| Observations | 64 | 36 | 28 |
| Adjusted R² | -0.01 | 0.10 | -0.04 |

*Note:* *p<0.1; **p<0.05; ***p<0.01

Note: The first column regresses monetary policy shocks on TFSI measured ahead of the monetary policy announcement, according to the specification: $MP\ S2_t\ 0 = a + \beta\ TFSI_{[t-1](4pm-1:59pm)} + \varepsilon_t$. The second and third columns restrict the sample to consider only tightening and easing shocks respectively. JK and Bauer and Swanson shocks are available for the sample period Sept. 2011- Dec. 2019. MAR shocks are for the 2-year on-the-run Treasury yield and constructed for the sample period Sept. 2011 - Dec. 2022. Note that MAR shocks beyond 2018 are currently confidential and available only within the Federal Reserve System. The tables reports OLS standard errors for all coefficient estimates (in parentheses).

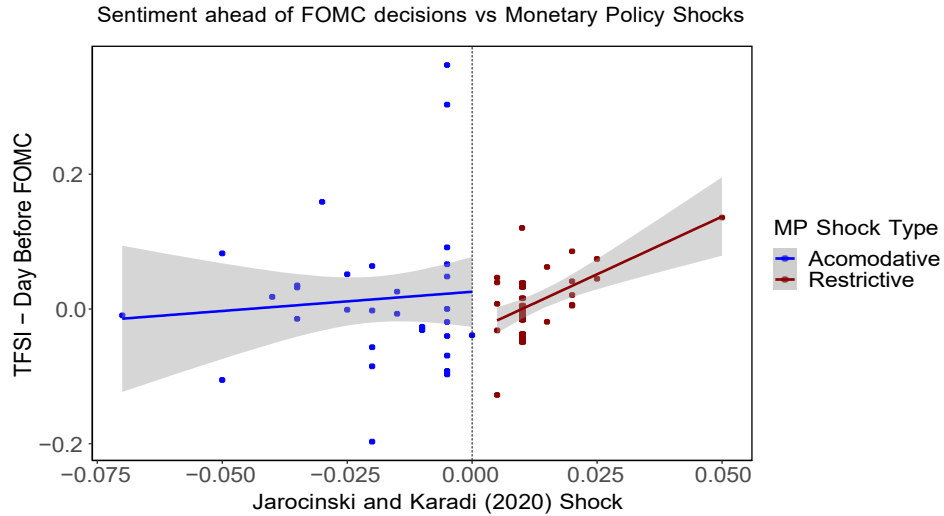

Sentiment ahead of FOMC decisions vs Monetary Policy Shocks

Source: Adams, Ajello, Silva, Vazquez-Grande, FEDS Paper #34, May 2023

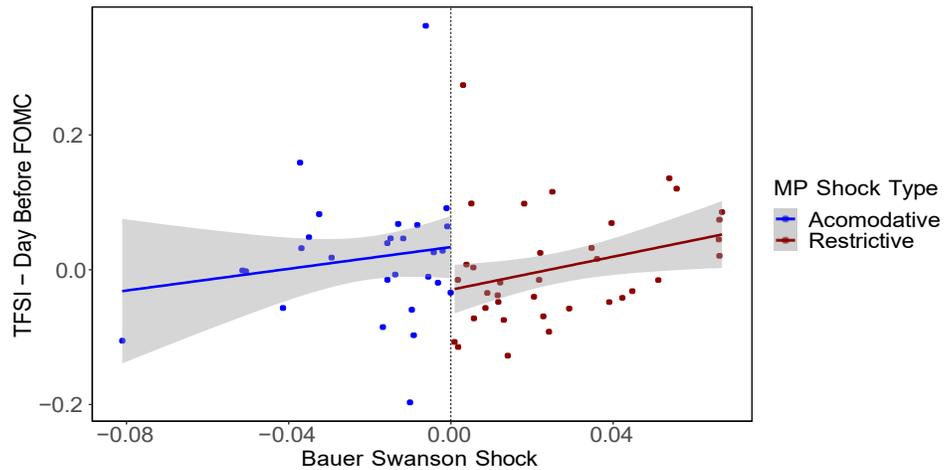

Note: The top panel displays the Jarocinski and Karadi (2020) monetary policy shocks against the TFSI computed between 4pm the day before and 2pm the day of the monetary policy decision (sample Sept. 2011 - Dec. 2019). The bottom panel displays the Bauer and Swanson (2022) monetary policy shocks against the TFSI computed between 4pm the day before and 2pm the day of the monetary policy decision (sample Sept. 2011 - Dec. 2019).
Source: TFSI: Authors' calculation based on Twitter enterprise-level API data; Jarocinski and Karadi (2020); Bauer and Swanson (2022)

significant predictor of sentiment after the policy announcement, independently of the sign of the monetary policy shock.[6] Our findings suggest that easing monetary policy shocks have no effect on the TFSI, while Twitter financial sentiment deteriorates both ahead and after a tightening monetary policy shock.

# 5 Conclusions

We build a real-time Financial Sentiment Index applying sentiment analysis to a query of tweets related to financial- and credit-market dictionaries. We find that changes in users' engagement–rather than in average tweeted sentiment–drives most variation in the index, that Twitter financial sentiment correlates highly with market-based measures of financial conditions and that overnight Twitter sentiment helps predict daily stock market returns. We document that Fed-related tweets play a dominant role on FOMC days and that sentiment deteriorates ahead of unexpected contractionary changes in the monetary policy stance. We also document that sentiment deteriorates further with the size of unexpected monetary policy tightening, while the relationship between sentiment and monetary policy accommodation is muted.

---

[6]The statistical significance of these results is preserved after controlling for financial sentiment in media, as measured by Shapiro (2020), the returns of the SP500 and the level of the VIX index.

## Table 4: TFSI and Monetary Policy Shocks — Delayed Response

### MAR shocks

|  | \multicolumn{6}{c}{Dependent variable:} |
| --- | --- | --- | --- | --- | --- | --- |
|  | \multicolumn{6}{c}{$TFSI_{t_{(2pm)} \to t_{(4pm)}}$} |
|  | All | All | Tight | Tight | Ease | Ease |
| *MARShocks* | -0.01 | -0.06 | 0.45*** | 0.35** | -0.01 | 0.06 |
|  | (0.11) | (0.10) | (0.13) | (0.14) | (0.16) | (0.14) |
| $TFSI_{t-1_{(4pm)} \to t_{(1:59pm)}}$ |  | 0.47*** |  | 0.24* |  | 0.55*** |
|  |  | (0.09) |  | (0.14) |  | (0.14) |
| Constant | -0.01 | -0.01 | 0.00 | 0.00 | 0.00 | -0.00 |
|  | (0.11) | (0.10) | (0.12) | (0.12) | (0.16) | (0.13) |
| Observations | 93 | 93 | 52 | 52 | 41 | 41 |
| Adjusted $R^2$ | -0.01 | 0.20 | 0.19 | 0.22 | -0.03 | 0.26 |

### JK shocks

|  | \multicolumn{6}{c}{Dependent variable:} |
| --- | --- | --- | --- | --- | --- | --- |
|  | \multicolumn{6}{c}{$TFSI_{t_{(2pm)} \to t_{(4pm)}}$} |
|  | All | All | Tight | Tight | Ease | Ease |
| *JKShocks* | -0.06 | 0.04 | 0.41** | 0.30* | -0.30* | -0.09 |
|  | (0.13) | (0.10) | (0.17) | (0.17) | (0.17) | (0.14) |
| $TFSI_{t-1_{(4pm)} \to t_{(1:59pm)}}$ |  | 0.61*** |  | 0.35* |  | 0.67*** |
|  |  | (0.10) |  | (0.17) |  | (0.14) |
| Constant | 0.00 | 0.00 | -0.00 | -0.00 | -0.00 | -0.00 |
|  | (0.13) | (0.10) | (0.17) | (0.16) | (0.17) | (0.13) |
| Observations | 62 | 62 | 30 | 30 | 32 | 32 |
| Adjusted $R^2$ | -0.01 | 0.35 | 0.14 | 0.23 | 0.06 | 0.46 |

### BS shocks

|  | \multicolumn{6}{c}{Dependent variable:} |
| --- | --- | --- | --- | --- | --- | --- |
|  | \multicolumn{6}{c}{$TFSI_{t_{(2pm)} \to t_{(4pm)}}$} |
|  | All | All | Tight | Tight | Ease | Ease |
| *Bauer − SwansonShocks* | 0.26** | 0.23* | 0.39** | 0.35** | 0.01 | 0.07 |
|  | (0.13) | (0.12) | (0.16) | (0.14) | (0.20) | (0.18) |
| $TFSI_{t-1_{(4pm)} \to t_{(1:59pm)}}$ |  | 0.49*** |  | 0.45*** |  | 0.42** |
|  |  | (0.12) |  | (0.14) |  | (0.18) |
| Constant | 0.11 | 0.06 | 0.00 | -0.00 | 0.00 | 0.00 |
|  | (0.13) | (0.12) | (0.16) | (0.14) | (0.19) | (0.18) |
| Observations | 64 | 64 | 36 | 36 | 28 | 28 |
| Adjusted $R^2$ | 0.05 | 0.24 | 0.13 | 0.32 | -0.04 | 0.11 |

*Note:* \hfill *p<0.1; **p<0.05; ***p<0.01

Note: The first two columns regress the TFSI after the release of the FOMC statement between 2pm and 4pm on monetary policy shocks, over all FOMC meeting since 2011 (columns 1 and 2) and separated by positive (columns 3 and 4) and negative shocks (columns 5 and 6). The even-numbered columns include the value of the TFSI before the release of the FOMC statement. The tables report OLS standard errors for all coefficient estimates (in parentheses).

# A  Technical Appendix

## A.1  Background

Twitter data has multiple advantages over traditional data. First, it has the possibility to allow researchers to understand perceptions of financial conditions in the US by a broader set of participants compared to surveys of professional market participants and financial market prices. Twitter data enables researchers to analyze a mixed sample of expert and non-expert commentary. There are 35 million Twitter Daily Active Users generating more than 500 million tweets per day. Second, the Twitter platform can also act as an information transmission channel allowing institutions like the Federal Reserve to reach out to otherwise inaccessible sectors of the population. Third, given the character limit of each tweet, users are compelled to communicate in succinct messages to convey their ideas. Lastly, this real-time submission of comments allows for high frequency analysis. All these characteristics can make for a better understanding of the opinion of "Main Street" about financial conditions in the US.

However, certain challenges arise when working with Twitter data. The high volume of tweets available on different topics raises the challenge to obtain a pertinent set of tweets to the research question. Researchers use Google-like keyword searches to filter tweets, and, without careful consideration, can lead to overly broad or overly narrow data samples. In addition to this, most Twitter user accounts are not verified, allowing for robot-generated tweets inside the data. Furthermore, due to its relatively recent founding, in 2006, Twitter data analysis is only possible for most recent years. The number of tweets remains low until the user-base reached a critical mass. Given the challenges mentioned above, we designed a carefully curated query to download a feasible set of finance-related tweets using multiple Natural Language Processing tools, such as heuristic operators ('AND','OR') and semantic similarity between words.

## A.2  How to Construct the Query.

We extracted finance-related tweets by using a multi=step financially intrinsic query. First, we obtain a list of financial words from Calomiris and Mamaysky (2019) and Danielsson et al. (2020). Given the sheer volume of tweets we would download by searching for all the words in this list, we use Search Engine Optimization (SEO) Keyword Clustering to group keywords that are semantically similar into three groups. Similar processes are used by SEO engineers to ensure that search engines, like Google, show their websites amongst the most relevant results to a specific query or group of queries. This technique allows us to find financial tweets that are semantically relevant to our desired subject-matter. We assume that words which are semantically similar are interchangeable and can be separated by the Boolean operator, 'OR'. Second, we use a pre-trained machine learning model called Wiki2Vec to group our keywords by semantic similarity. This model, trained on the corpus of Wikipedia articles, converts words into computer-understandable vectors that allows to calculate word distance. We take these calculated similarities to determine optimal keyword groups, or clusters. We then generate three matrices based on 100, 300, and 500 vectors for each word to find similarity scores of each word compared to all other words in our dictionary. We highlight keyword pairs that have a similarity score higher than the matrix average to categorize word clusters. Based on our analysis and API limitations for monthly download of tweets, we chose to use three clusters.

We modify these clusters to allow for common phrases to be built from them. The first group, containing words like "bond", "security", "asset", or "currency", became the "noun/object" group. The second group is composed of "actors" or "subject" nouns, i.e., "company", "corporate", "market", or "Federal Reserve". The third group of words is comprised of "modifier" words or compounds, like "coupon", "term", "upgrade", and "default". The full set of words inside these groups can be found in Appendix A.

We eliminate from our query tweets that refer to advertising by excluding tweets that contain the specific phrases "social security" or "credit card". Furthermore, we do not download exact retweets to maximize our download quota. Regardless of this loss of retweets as individual observations, we obtain the retweet count and other "engagement" counts of each original tweet.

We extract the main informational set of each tweet, the text and date of publication, as well as their user metadata: username, verification status, and location, if available. We also have access to extended metadata related to each tweet including the aforementioned retweet count and other engagement counts, other tweets to which it might be in reply, and an assessment whether the tweet contains "possibly sensitive" or "NSFW" content. So far, we have processed 7.1 million tweets since the advent of Twitter, with an average increase of nearly 450,000 tweets per year. Due to the narrower scope of our search than general Twitter usage, there were few applicable results in the early days of Twitter. Fewer than 5,000 tweets were posted prior to 2009, with a little over 45,000 posted in 2009, and almost 150,000 posted in 2010. We choose to start our Twitter Financial Sentiment Index in 2011 given that more than 250,000 tweets were posted. The number of tweets that falls within our query has gone up on average ever since. The month with the highest number of tweets downloaded is October 2022, with about 188,000 tweets, while the month with the lowest number of tweets is February 2011, with around 10,000 tweets.

## A.3   Processing the Raw Data

Generating the Twitter Financial Sentiment Index requires a few steps after obtaining the data. First, we preprocess the text of each tweet by removing "@" tags, excess white space, hyperlinks, and replace common ascii plain text with the appropriate special character, i.e. "&" becomes "&". We also eliminate from our sample of tweets any tweet that refers to cryptocurrency and decentralized financial assets by removing tweets that contain the text strings "crypto","token","NFT","inu","shiba", and "defi". In addition to this initial text cleaning, we separate tweets with exact duplicate text into another data set to later be included in engagement counts on each unique tweet. Second, we shift the time zone of tweets to reflect Eastern Standard Time (EST) instead of the default Coordinated Universal Time (UTC) for easier comparison to United States events. Third, we calculate the sentiment values of each tweet in our sample using the Bidirectional Encoder Representations from Transformers (BERT) sentiment. Fourth, we set to zero the sentiment of tweets that contain values between -.1 and .1. Lastly, we obtain an index by taking the simple mean of all the remaining tweets from 2011 to 2022 by different frequencies– daily, weekly, and monthly.

A notable characteristic of our sample is the presence of tweets with the same text content and sentiment score. We address these tweets as "manual" retweets. Though our query explicitly excludes retweets from appearing in our search, this only prevents retweets that were made by clicking the retweet button. Oftentimes a manual retweet is done by bots, or by users sharing the exact same content from a non-twitter website, like news articles, editorials, and blog posts. To control for this, we remove tweets if the text is duplicated in a previous tweet, keeping the first incidence. There are 2M manual retweets in our sample. After removing, our sample size is 5.1M tweets. We then proceed to calculate the sentiment of each tweet.

## A.4 FinBERT

We use Google's Bidirectional Encoder Representations from Transformers (BERT) as our starting point to analyze the sentiment of tweets. BERT is a state-of-the-art pre-trained machine learning model capable of understanding sentences alongside the context in which they are being applied. BERT is pre-trained on the Toronto BookCorpus (containing 800M words) and Wikipedia articles (containing 2.5B words). BERT converts words into vectors, and reads the text bidirectionally to classify sentences given the context in which words are being used. This unique ability to understand contextual representation, and doing so in both directions of the text allows BERT to significantly outperforms other machine-learning-based and dictionary-based models in tasks like text prediction and sentiment calculation. Furthermore, it can be pre-trained further and then fine-tuned to better understand a desired context, like financial jargon.

We use the model FinBERT as our baseline for sentiment scoring. FinBERT, from Araci (2019), is a refined version of BERT that is designed to understand text in the context of Financial sentiment. FinBERT is pre-trained using a large corpus of financial texts and fine-tuned with a dictionary of financial words and phrases from Malo et al. (2014). One caveat of FinBERT is that it was pre-trained using longer texts, so it splits sentences individually and then calculates sentiment on each one of them. Given the context of tweets can be better understood as a whole, rather than separated by sentences, we replace full sentence stops, ".", with a semi-colon, ";", to "desentencize" our text before calculating sentiment values on each of our tweets.

FinBERT produces five sentiment values. Three values represent the probabilities that the text is either positive, negative, or neutral. FinBERT also calculates a compound score as the positive probability minus the negative probability. Lastly, FinBERT provides trinary sentiment prediction which is based on the highest of the three probabilities. We drop tweets that are classified as neutral in this prediction (neutral probability is highest). Then, we obtain a sentiment score for each tweet in our sample.

We calculate our Twitter Financial Sentiment Index as the negative of the average the sentiment of all tweets by a given frequency, either daily, weekly, or monthly. The values of this measure range from -1 (extremely positive) to 1 (extremely negative). The monthly index has a mean of -0.09, and a standard deviation of 0.1.

## A.5 Valence Aware Dictionary for Sentiment Reasoning

We also analyze a sentiment with a second model as a robustness check to our baseline BERT measure. We derive sentiment values from the Valence Aware Dictionary for sEntiment Reasoning (VADER) model. This dictionary-based sentiment scorer is specifically designed for analyzing sentiment on social media platforms, which tend to be shorter, more cryptic messages than the average texts analyzed. This dictionary is capable of understanding phrases, emoticons, and acronyms like 'XD" and "LOL". Furthermore, VADER can comprehend semantic modifiers from heuristic rules that other dictionaries lack. This key feature of VADER allows the calculation of the magnitude of sentiment, not only its polarity. For example, using VADER, the phrase "really bad" produces a more negative sentiment value than the single word "bad", while the phrase "not bad" produces a positive sentiment value.

Given VADER is a dictionary-based sentiment scorer, we seek to prevent selection bias by neutralizing the sentiment score of all the words from our constructed query. From all the words used in the query, only the words 'asset', 'credit', 'cut', 'debt', 'interest', 'low', 'pay', 'security', 'share', and 'treasury', and their plurals contained a sentiment valence other than zero. We then calculate the positive, neutral, negative, and compound sentiment values of each tweet in our sample.

Once the sentiment score is calculated, we filter out the tweets with an absolute compound score lower than 0.1. From our sample of 5.1M tweets, we removed 1.7M neutral tweets under this threshold. We argue that this tweets that show neutral sentiment are purely informational and serves the purpose of transmitting information to twitter users, rather than expressing sentiment. It is important to note that the distribution of informational tweets overtime is not uniform. From 2011 to 2017, informational tweets formed roughly 50% of the share of all tweets while the share of zero-valued tweets decreased to 25% from 2018 onward. This drastic change in the share of information tweets also came with an increase in the share of positive and negative tweets, from roughly 25% of the share overtime for both, to 50% and 25% of the share overtime respectively. This change in the composition of the sentiment-charged tweets in the sample is due to Twitter's executive decision to increase the character length limit of each tweet that took effect in December 2017. The increase in character limit from 140 to 280 increased the number of average words per tweets from 15.3 to 32.2. This gave more space for demonstrating more sentiment in each tweet by increasing the probability of writing more sentiment charged words. This is also an effect of VADER. We see, in general, that the longer a text that is provided to VADER, the more polarized the sentiment score becomes.

When comparing to the BERT-based TFSI, we see a noticeable change in the shares of positive, negative, or neutral tweets when the character limit was increased in late 2017. This is likely due to finBERT's superior ability to understand context and account for differences in the number of sentiment-laden words. These shares are relatively consistent over time with roughly 50% of tweets being classified as neutral and the remaining 50% a relatively even split between positive and negative. The linear correlation between the BERT-based and VADER-based indexes is 0.68.

## A.6 Twitter-based Financial Sentiment Index

We construct our Financial Sentiment Index by obtaining the simple mean of the compound sentiment score for all remaining after pre-processing steps take place. We do this at weekly, and monthly frequencies. Given the higher level of variance at higher frequencies, we use the 7-day moving average of TFSI values to obtain the daily TFSI.

We found our Twitter Financial Sentiment monthly index is highly correlated to other measures of financial conditions, like the Financing Conditions Index for Non-Financial Corporations, produced at the Federal Reserve Board. It has a correlation of 0.79 with the FCINFC. Also, it responds congruently to negative and positive shocks. The average sentiment across time is 0.12 and the standard deviation is 0.07. The volatility of earlier years in the index is higher, arguably due to the lack of a meaningful number of observations per month. There are two significant lows since the character limit increase, one in September 2019, and the other in March 2020.

## A.7 Fed-worded Tweets on FOMC days

We also seek to understand the transmission of information on the Federal Reserve via tweets around events directly related to it. To do this, we analyze the traffic of Tweets that are related to the Federal Reserve around the Federal Open Markets Committee events. First, we generated a subsample of tweets that contained the text string 'Fed','Reserve','monetary', 'Powell', and 'Yellen' respective of their time as Federal Reserve Chairs. The average volume of tweets containing these strings is 65 per day. We found that, on average, this volume increased to 231 tweets 24 hours after an FOMC event took place. This increase in volume of tweets is 150 tweets above the 3rd quartile of the volume of all Fed-worded tweets overtime. This indicates that Twitter is used as a transmission channel for the information conveyed in FOMC events. We then proceed to analyze the relationship of the frequency of this transmission and monetary policy shocks.

## A.8 Back-testing a Trading Strategy

In order to assess a practical use of the TFSI, we build on the index to construct trading signals and back-test them in our sample. As positive overnight values of the TFSI lead to negative open-to-close returns on the S&P500 index at the subsequent trading day (as shown in Table 1), we consider two trading strategies. First, a negative value of overnight TFSI would lead us to go long the S&P500, buying at open values and selling that same day at closing values, while positive overnight values of the TFSI would lead us in turn to short the S&P500 over trading hours. This trading strategy—that we call "TFSI unrestricted"—is appealing for its simplicity as it does not depend on estimates of a model to be implemented, it also minimizes data-mining. The second strategy, considers that the TFSI measures sentiment with error, and therefore values close to 0 should be taken as less of a signal that those farther away from 0. Predicated on this insight, we design a trading strategy that shorts the S&P at opening when the TFSI is above the 0.12, and goes long the S&P500 when the TFSI is below -.12, the rest of days the there would be no trade—we call this strategy "TFSI threshold". The choice of this threshold is not very restrictive of the number of trades, leading to trading 84% of days after CLI, and 87% of days over the full sample. As a benchmark we compare our returns to the daily returns of the S&P 500 index.[7] We abstract from considering transaction costs.

Table 5 displays selected return-distribution statistics to assess the performance of the TFSI strategies and the returns of the S&P500 over the post-CLI and full samples, on the left and right panels respectively. We compute (annualized) average returns, and (annualized) standard deviation, information ratio (the ratio of annualized average returns and annualized standard deviation), skewness and kurtosis. We also present measures of Value-at-Risk (VaR) and expected shortfall (ES) to compare the performance of strategies during tail events. We compute the VaR and ES measures using the historical distributions of returns over the two samples, at the 95% level.

---

[7]As an alternative benchmark we have also used a trading strategy that goes long the S&P500 index over US trading hours, buying at the open valuation and closing the position at the closing valuation. Using this strategy as a benchmark does not change the takeaways from this exercise.

Table 5: Return Moments across Trading Strategies

|  | Post-CLI Sample | | | Full Sample | | |
| --- | --- | --- | --- | --- | --- | --- |
|  | *TFSI* threshold | *TFSI* unrestricted | *SP*500 | *TFSI* threshold | *TFSI* unrestricted | *SP*500 |
| Avg. Ret (ann) | 11.2% | 8.0% | 8.3% | 9.6% | 8.0% | 10.5% |
| St. Dev. Ret (ann) | 15.1% | 16.1% | 21.4% | 13.5% | 14.3% | 17.6% |
| Inf. Ratio | 0.75 | 0.50 | 0.39 | 0.71 | 0.56 | 0.60 |
| Skewness | −0.14 | −0.14 | −0.80 | −0.06 | −0.08 | −0.73 |
| Kurtosis | 6.31 | 4.46 | 13.33 | 5.77 | 4.37 | 14.65 |
| Value-at-Risk (95%) | −1.3% | −1.5% | −2.1% | −1.3% | −1.4% | −1.3% |
| Exp. Shortfall (95%) | −2.3% | −2.4% | −3.4% | −2.0% | −2.1% | −2.7% |

Table 5 shows the reasons of the out-performance of the TFSI strategies relative to the S&P500: the TFSI strategies have a lower volatility and a higher value of skewness in the distribution of returns—that is, the TFSI strategies (column 1 and 2 in both panels) produce less volatile and a more symmetric distributions of returns relative to the S&P 500 daily returns (column 3 in both panels). We interpret this as the TFSI helping to forecast the sign of the return of the S&P500 over US business hours, on average, reducing Value-at-Risk and the expected shortfall relative to the the S&P500 returns. In the post-CLI sample, when the TFSI is able to capture a stronger signal, both TFSI strategies show a significantly higher information ratio relative to the market. For the TFSI threshold strategy, this extends across samples as well with higher average returns and lower volatility, and lower measures of tail-risk, making it significantly preferred to the benchmark.

# B  Word Clusters

Table 6: Word Clusters

| Group 1 | Group 2 | Group 3 |
|---|---|---|
| Bond | Corporate | Coupon |
| Debt | Company | Interest |
| Security | Subsidiary | Rate |
| Credit | Market | IPO |
| Loan | Municipal | Term |
| Mortgage | Sovereign | Liquidity |
| Portfolio | Program | Yield |
| Pension | Market | Downgrade |
| Federal Funds | Federal Reserve | Outstanding |
| Leverage | Collateral | Repayment |
| Financing | Credit Agency | Default |
| Rent | Sovereign | Initial Public Offering |
| Portfolio | Credit Agency | Lending |
| Asset | Program | Return on |
| Pension | Federal | Upgrade |
| Facility | Federal Reserve | |
| CPFF | Money Markets | |
| PDCF | Collateral | |
| MMMFLF | Junk | |
| PMCCF | High Yield | |
| MLF | Investment Grade | |
| MSCP | HY | |
| Fund | IG | |
| Currency | | |
| Debenture | | |
| Leverage | | |
| Cash | | |
| Finance | | |
| Financial Leverage | | |